# Quick and repeatable shear modulus measurement based on torsional resonance and torsional wave propagation using a piezoelectric torsional transducer

Mingyu Xie[1,2], Qiang Huan[1,2], Faxin Li[1,2,3,a]

[1] LTCS and College of Engineering, Peking University, Beijing 100871, China
[2] Center for Applied Physics and Technology, Peking University, Beijing, China
[3] Beijing Key Laboratory of Magnetoelectric Materials and Devices, Peking University, Beijing, China

**Abstract**

Shear modulus is one of the fundamental mechanical properties of materials, while its quick and accurate measurement is still a challenge. Here we proposed two methods for shear modulus measurement based on torsional resonance and torsional wave propagation using a same piezoelectric torsional transducer. Firstly, the torsional transducer was introduced which consists of two thickness poled, thickness shear ($d_{15}$) piezoelectric half-rings. Secondly, the equivalent circuit of the transducer-cylindrical specimen system is derived and the shear modulus can be explicitly obtained using the torsional resonance frequency. The internal friction can also be obtained, which is calculated by using an approximate formula. Then, the principle of shear modulus and internal friction measurement based on torsional wave propagation were presented. Finally, shear modulus and internal friction measurement on four materials including 1045 steel, aluminum, quartz glass and PMMA, were conducted. Results indicate that the measured shear moduli using these two methods are very close to each other, and consistent with the reference values in literatures. The resonance method is quite convenient and highly repeatable, but is typically not suitable for long specimens where the first torsional resonance may not be visible. The wave propagation method is especially suitable for long specimens and high frequency measurement is suggested. The internal frictions measured by these two methods were also close to each other, and the simple wave attenuation method is suggested. The two shear modulus measurement methods proposed in this work are quite reliable and can be widely used in near

---

[a] Author to whom all correspondence should be addressed, Email: lifaxin@pku.edu.cn

future.

1. Introduction

Elastic moduli are the fundamental material properties and of great importance in engineering and science. For isotropic materials, there only exist two independent elastic moduli, i.e., the Young's modulus (E) and the shear modulus (G). Currently there are quite a few methods that can be used to measure the Young's modulus and shear modulus. It is well known that the static moduli can be measured by using the applied quasi-static tension (compression) and torsion. If not mandatory, the quasi-static methods are usually not suggested as the testing errors are typically large which can reach 5% or even larger. Dynamic testing methods are generally of good accuracy and most methods are valid for both Young's modulus and shear modulus measurement. In a famous interlaboratory testing program in 1980s,[1] six dynamic testing methods were used, i.e., 1) free-free beam resonance;[2] 2) Impulse excitation technique;[3] 3) Ultrasonic wave velocity method;[4] 4) Piezoelectric ultrasonic composite oscillator technique (PUCOT);[5] 5) Ultrasonic pulse spectroscopy;[6] 6) Magnetically excited resonance method.[7] The measured Young's moduli of two Inconel alloys using different methods agreed well with each other, with the maximum relative error of only 1.6%. However, in that program, measurement on the shear modulus were not reported or not conducted.

Actually, shear modulus is of special importance in solid state matters. For example, gradual variations of shear modulus had been observed in polymers, glass, metal glass, etc., during phase transitions (or glass transitions).[8-10] Thus, accurate measurement of shear modulus can be used to monitor these complicated physics process. Besides above-mentioned methods that can be used to measure both Young's modulus and shear modulus,[2-7] there is also a special method only applicable to shear modulus measurement, i.e., the torsion pendulum.[11] It should be noted that the existing methods for shear modulus measurement were either somewhat inaccurate or not easy to conduct. The free-free beam method,[2] which measured the shear modulus based on the torsional resonance of a rectangular bar-shaped specimen, is strongly dependent on the support location and support manner. Furthermore, to make the torsional resonance not very high (typically below 30kHz) to avoid the excitation difficulty, the length of the bar cannot be very short (typically larger than 80mm) and the thickness cannot be large (typically less than 3mm for metals), thus the fabrication errors on the thickness is difficult to control and this would introduce significant errors in shear modulus measurement. The impulse excitation technique,[3] which excites the specimen by a light mechanical impulse and analyzes the resultant transient vibration, required a complicated circuit and algorithm to extract the fundamental frequency thus the modulus. For the ultrasonic wave velocity method,[4] the shear modulus measurement were conducted on a bar or rectangular specimen. Typically a shear-mode transducer were bonded or coupled onto the lateral face of the specimen and the round-trip travelling time of high frequency pulses (typically above 5MHz) in the specimen were extracted to calculate the shear wave speed thus the shear modulus. Note that as the thickness of the specimen is on the same order with the wavelength, actually complicated guided wave propagates in the specimen. Thus, extraction of phase velocity instead of the ground velocity is not straightforward,[12] typically based on the pulse echo superposition method[13] or the pulse echo overlap method.[14, 15] The PUCOT[5, 16] is typically only used for measurement of Young's modulus and mechanical damping. Shear modulus measurement using PUCOT were never reported although torsional oscillation were also excited for damping measurement.[17] The ultrasonic pulse spectroscopy[6] and the magnetically excited resonance method[7] were not commonly used due to their complicated testing components and/or algorithm. The torsion pendulum method is typically only applicable to thin wires (bars) or soft materials,[11, 18, 19] thus it cannot act as a general method for shear modulus measurement and the testing error is relatively large, typically of several percent.[19]

As indicated in the ASTM standard,[20] theoretically shear modulus measurement using a cylinder specimen is both simpler and more accurate than that using a rectangular bar. However, experimental difficulties in exciting torsional resonance usually preclude its use in determining the shear modulus. Johnson et al excited the resonant torsional modes in a cylindrical aluminum rod using electromagnetic acoustic transducers (EMAT) consisting of 24 rows of magnets and meander coils and measured the shear wave velocity with very good consistence.[21] Note that the EMAT method is only applicable to metals and the complicated structure of the EMAT makes this method not suitable for practical shear modulus measurement.

In this work, we firstly proposed a piezoelectric torsional transducer based on two thickness-poled, thickness-shear ($d_{15}$) PZT half-rings. Then, the torsional transducer was bonded on one end of a cylinder specimen to excite torsional vibration and torsional guided waves. Equivalent circuit model[22] was used to analyze the transducer-specimen composite system and the shear modulus of the specimen can be explicitly obtained by measuring the electromechanical resonance of the transducer-specimen system using an impedance analyzer. The non-dispersive fundamental torsional wave T(0,1) can be excited by using this torsional transducer. Thus measurement of the shear wave velocity and then the shear modulus is straightforward. Formulas of the internal friction measurement using these two methods were also presented. Then, both the shear modulus and internal friction measurement were conducted on four types of materials including 1045 steel, aluminum, quartz glass and PMMA. The measurement results were compared with each other and consistent with the reference values. This work provides a quick and highly repeatable solution to the shear modulus measurement on cylinder samples, which can also be used to study the shear modulus variations during or after mechanical loading (tension/compression, torsion, etc.) where circular cross-section specimen are commonly employed.

## 2. Principle of shear modulus measurement based on a piezoelectric torsional transducer

### 2.1 The thickness-poled piezoelectric torsional transducer

As mentioned above, more accurate shear modulus measurement can be realized by excitation of torsional resonance in a cylinder sample while the structure of the EMAT to excite torsional vibration is too complicated for practical testing.[19] A piezoelectric torsional transducer should excite torsional vibration of a cylinder sample easily in the contact mode. However, the well-known shear mode quartz oscillator is not a good torsional transducer since its properties are not uniform along the circumferential direction.[17] The currently available Langevin-type piezoelectric torsional transducer is rather difficult to fabricate since it is circumferentially poled after many steps.[23] Note that it is almost impossible to make the synthetic circumferential polarization uniform, especially when its diameter gets smaller.

Here we proposed a novel piezoelectric torsional transducer consisting of two thickness-poled piezoelectric half-rings, which is very easy to fabricate. As shown in Fig.1, a PZT ring (with the outer diameter D, inner diameter d, and thickness h) is firstly poled along the thickness direction. After poling, the ring is evenly cut into two half-rings. Then, the top/bottom electrodes were removed and lateral electrodes were spread on the cutting faces of the two half-rings. Finally, the two half-rings were bonded together using conductive epoxy with the poling directions opposite to each other. Thin copper wires were inserted between the bonding faces for electrical excitation/reception.

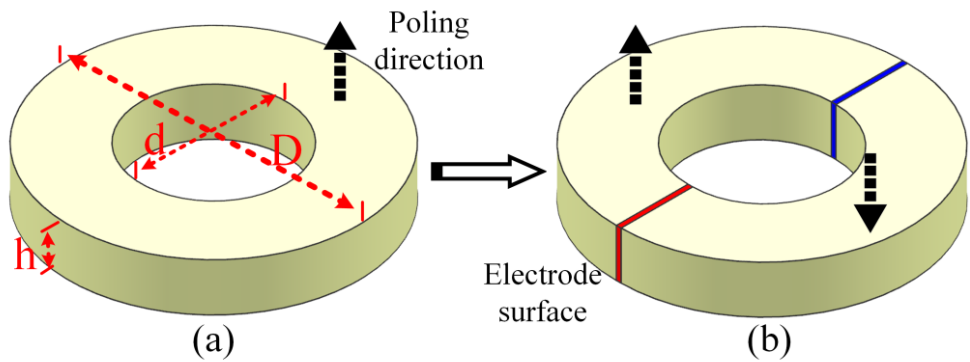

Fig.1 The piezoelectric torsional transducer consisting of two thickness-poled piezoelectric half-rings.

## 2.2 Equivalent-circuit model for the piezoelectric torsional transducer

### 2.2.1 Theory

We firstly analyze the vibration mode of the thickness-poled piezoelectric half-ring. As shown in Fig.1b, when an electric field is circumferentially applied to the thickness-poled piezoelectric half-ring, there only exists the circumferential displacement in the half-ring, i.e., both the axial and radial displacements should be zero. Furthermore, the electric field should be uniform along the circumferential direction. Thus, we have

$$\begin{cases} u_\theta(r,z,t) = r\Theta_P(z,t) \\ E_\theta = \dfrac{V_\theta}{\pi r} \end{cases} \tag{1}$$

Here $u_\theta$ is the circumferential displacement, $\Theta_P$ is the angular displacement, $V_\theta$ is the applied voltage and $E_\theta$ is the electric field.

Then the only non-zero shear strain of the half-ring can be expressed as

$$\gamma_{\theta z} = r\frac{\partial \Theta_P(z,t)}{\partial z} \tag{2}$$

The piezoelectric constitutive equations can be written as

$$\begin{cases} \gamma_{\theta z} = s_{44}^E \sigma_{\theta z} + d_{15} E_\theta \\ D_\theta = d_{15}\sigma_{\theta z} + P_{11}^\sigma E_\theta \end{cases} \tag{3}$$

where

$$\begin{cases} s_{44}^E = s_{44}'^E - j s_{44}''^E \\ p_{11}^\sigma = p_{11}'^\sigma - j p_{11}''^\sigma \\ \dfrac{1}{Q_{mP}} = \dfrac{s_{44}''^E}{s_{44}'^E} \\ \dfrac{1}{Q_e} = \dfrac{p_{11}''^\sigma}{p_{11}'^\sigma} \end{cases} \tag{4}$$

Where $s_{44}^E$ is the elastic coefficient in short circuit, $d_{15}$ is the piezoelectric coefficient, $D_\theta$ is the electric displacement along the circumferential direction, $P_{11}^\sigma$ is the stress-free dielectric constant. $Q_{mP}$ is the mechanical quality factor of the PZT and $Q_e$ is the electric quality factor of the PZT.

The equilibrium equation is

$$\frac{\partial \sigma_{\theta z}}{\partial z} = \rho_P \frac{\partial^2 u_\theta}{\partial t^2} \tag{5}$$

where $\rho_P$ is the mass density of the piezoelectric half-ring.

By substituting Eq.(3a) into Eq.(5) and bearing in mind that the circumferential electric field is constant along the thickness, we can get the vibration equation of the piezoelectric half-ring to be:

$$\frac{\partial^2 \Theta_P(z,t)}{\partial t^2} = c_P^2 \frac{\partial^2 \Theta_P(z,t)}{\partial z^2} \tag{6}$$

where

$$\begin{cases} c_P = \sqrt{\dfrac{1}{s_{44}^E \rho_P}} = c_P' + jc_P'' \\ c_P' \approx \dfrac{1}{\sqrt{\rho_P s_{44}'^E}} \\ c_P'' \approx \dfrac{1}{\sqrt{\rho_P s_{44}'^E}} \dfrac{1}{2Q_{mP}} \end{cases} \quad (7)$$

The approximate solution of Eq.(6) is

$$\Theta_P(z,t) = (A_1 \cos k_P z + B_1 \sin k_P z) e^{j\omega_P t} \quad (8)$$

where

$$k_P = \frac{\omega_P}{c_P'} - j\frac{1}{2Q_{mP}}\frac{\omega_P}{c_P'} \quad (9)$$

Here $c_P$ is the shear wave velocity of the piezoelectric ceramics in short circuit, $\omega_P$ is the angular frequency. $k_P$ is the wavenumber whose real part determines the speed of sound and the imaginary part determines the attenuation of the wave amplitude.

For the general solution in Eq.(8), the boundary conditions of the velocity is

$$\begin{cases} \dot{\Theta}_P \big|_{z=0} = U_1 \\ \dot{\Theta}_P \big|_{z=h} = -U_2 \end{cases} \quad (10)$$

Then we can get

$$\begin{cases} A_1 = \dfrac{U_1}{j\omega_P} e^{-j\omega_P t} \\ B_1 = -\dfrac{1}{j\omega_P}\left(\dfrac{U_1}{\tan k_P h} + \dfrac{U_2}{\sin k_P h}\right) e^{-j\omega_P t} \end{cases} \quad (11)$$

Based on the mechanical boundary conditions:

$$\begin{cases} -F_1^* = \int r\sigma_{\theta z}\big|_{z=0} dA \\ -F_2^* = \int r\sigma_{\theta z}\big|_{z=h} dA \end{cases} \quad (12)$$

The transport equations for the vibrations of the piezoelectric torsional half-ring can be obtained to be:

$$\begin{cases} F_1^* = \left( Z_P^* j \tan \dfrac{k_P h}{2} + \dfrac{Z_P^*}{j \sin k_P h} \right) U_1 + \dfrac{Z_P^*}{j \sin k_P h} U_2 + N^* V_\theta \\ F_2^* = \dfrac{Z_P^*}{j \sin k_P h} U_1 + \left( Z_P^* j \tan \dfrac{k_P h}{2} + \dfrac{Z_P^*}{j \sin k_P h} \right) U_2 + N^* V_\theta \\ I^* = -N^* U_1 - N^* U_2 + j\omega_P C_0^* V_\theta \end{cases} \quad (13)$$

Where $Z_P^*$ is the impedance of the piezoelectric torsional half-ring, $N^*$ is the electromechanical conversion coefficient of the piezoelectric torsional half-ring, $C_0^*$ is the static capacitance, $P_{11}^\varepsilon$ is the constrained dielectric constant, and

$$\begin{cases} Z_P^* = \dfrac{k_P \pi (D^4 - d^4)}{64 s_{44}^E \omega_P} \\ N^* = \dfrac{d_{15}(D^2 - d^2)}{8 s_{44}^E} \\ C_0^* = \dfrac{h}{\pi} P_{11}^\varepsilon \ln \dfrac{D}{d} \\ P_{11}^\varepsilon = P_{11}^\sigma - \dfrac{d_{15}^2}{s_{44}^E} \end{cases} \quad (14)$$

Now let us analyze the vibration mode of the piezoelectric torsional ring consisting of two half-rings with the opposite polarization. It can be easily deduced that the transport equation will keep the same form as Eq.(13) but the related coefficients will be doubled, i.e., $N=2N^*$; $Z_P=2Z_P^*$; $C_0=2C_0^*$ will replace the coefficients $N^*$, $Z_P^*$ and $C_0^*$ respectively and $F_1 = 2F_1^*$; $F_2 = 2F_2^*$; $I = 2I^*$. Then, the general equivalent circuit of the torsional piezoelectric ring can be obtained, as plotted in Fig.2(a) where the mechanical boundary conditions for the up/bottom faces are not prescribed. For the usual case, both the up and bottom faces of the ring is stress-free and the corresponding equivalent circuit turns to be Fig.2(b). According to Fig.2(b), we can get the expression of the admittance to be:

$$Y = jC_0 \omega_P + \dfrac{N^2}{\dfrac{-Z_P j}{2} \cot \dfrac{k_P h}{2}} \quad (15)$$

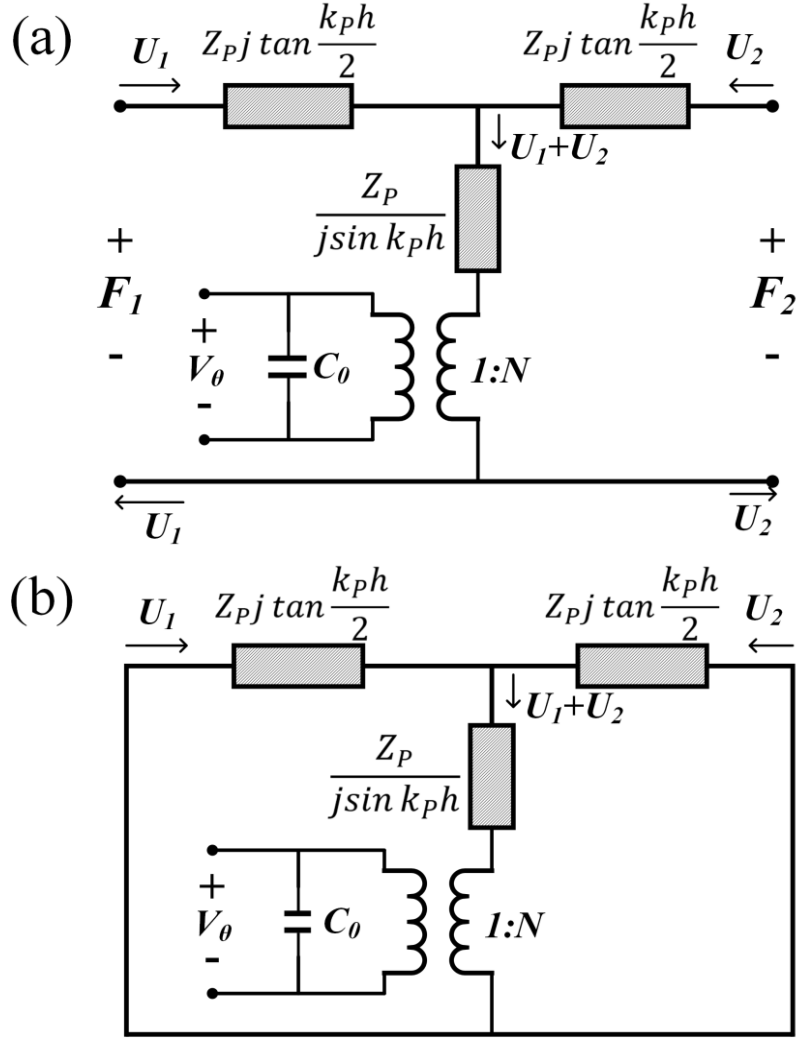

Fig.2  The equivalent circuit of the piezoelectric torsional transducer consisting of two thickness-poled half-ring. (a) General case; (b) The case with the up/bottom faces stress-free.

**2.2.2 Experimental verification**

To verify the equivalent circuit model of the piezoelectric torsional transducer, we measured the frequency dependent admittance curve of the transducer and the results were shown in Fig.3 where the theoretical results calculate from Eq. (15) were also presented for comparison. The size of the piezoelectric torsional transducer was measured by using a micrometer caliper and the results are: outer diameter of 12.00mm±0.06mm, inner diameter of 5.40mm±0.04mm, and thickness of 1.92mm±0.02mm. The material constants of the piezoelectric ring were provided by the manufacturer and listed as follows: the density $\rho_P = 7500\text{kg/m}^3$, elastic constant $s_{44}'^E = 43.5 \times 10^{-12}\text{m}^2/\text{N}$, piezoelectric constant $d_{15} = 741 \times 10^{-12}\text{C/N}$, relative dielectric constant $P_{11}'^\sigma/\varepsilon_0 = 3130$ and $\varepsilon_0$ is the vacuum permittivity. Note that the electrical quality factor has no effect on the sharpness of the resonance peak. It only affects the slope of the whole curve, so we use the parameters $Q_e$=50 supplied by the manufacturer. By fitting the testing results, the mechanical quality factor is extracted to be $Q_{mP}$=18. This is not consistent with the manufacturer's value of 65, which may be due to the fact that the quality factor is not only related to the material, but also to the vibration mode.

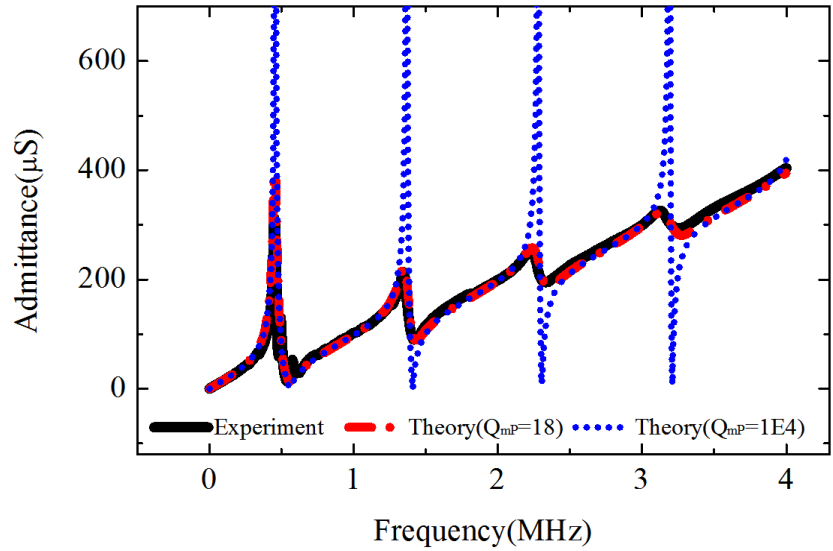

Fig.3 The measured and theoretical (with different mechanical quality factors) admittance curves of the piezoelectric torsional transducer

It can be seen from Fig.3 that overall the theoretical curves with $Q_{mP}$=18 can fit quite well with the measured curves, which indicated the validity of the equivalent circuit model.

**2.3 Shear modulus measurement based on electromechanical torsional resonance**

If the above-mentioned piezoelectric torsional transducer is perfectly bonded onto a cylindrical specimen with its diameter equal to the outer diameter of the transducer, as shown in Fig.4, it is expected that torsional vibration can be excited in the cylindrical specimen. Since the torsional resonance frequency of the cylindrical specimen (typically ~several 10kHz for metals) is much lower than the corresponding torsional resonance frequency of the piezoelectric transducer (~460kHz), the resonance frequency of the transducer-specimen composite system should be very close to the specimen's resonance frequency as the specimen's mass is typically much larger than that of the transducer.

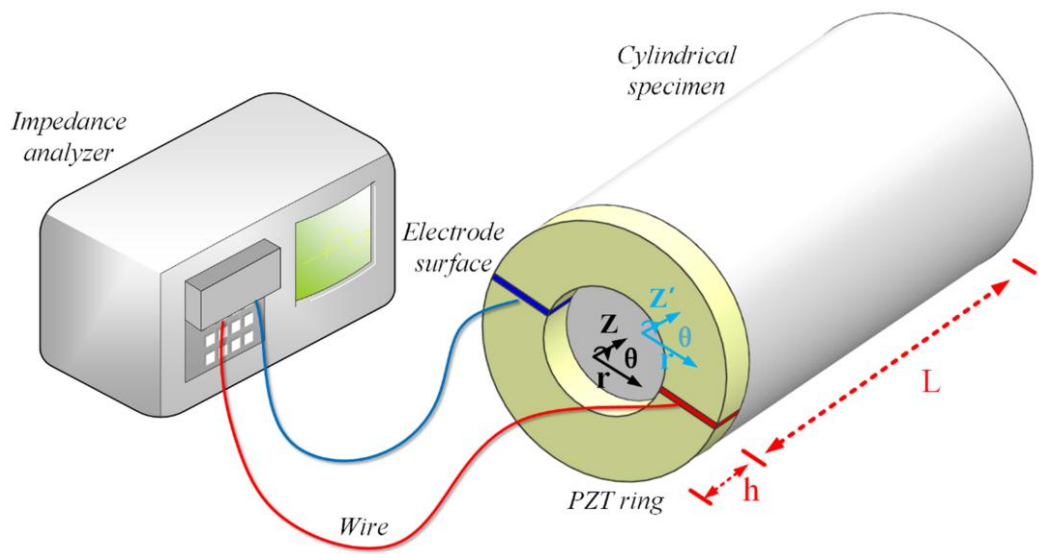

Fig.4 Testing setup for shear modulus measurement based on electromechanical torsional resonance using a piezoelectric torsional transducer

### 2.3.1 Equivalent circuit model of the transducer-specimen composite system

Considering the internal friction, the shear modulus of the specimen can be written as

$$\begin{cases} G = G' + jG'' \\ \dfrac{1}{Q_{mM}} = \dfrac{G''}{G'} \end{cases} \quad (16)$$

For a single cylindrical specimen, the equilibrium equation for its torsional vibration can be expressed as

$$\frac{\partial^2 \Theta_M(z',t)}{\partial t^2} = c_M^2 \frac{\partial^2 \Theta_M(z',t)}{\partial z'^2} \quad (17)$$

The general solution for Eq.(17) is:

$$\Theta_M(z',t) = \left(A_2 \cos k_M z' + B_2 \sin k_M z'\right) e^{j\omega_M t} \quad (18)$$

where

$$\begin{cases} c_M = \sqrt{\dfrac{G}{\rho_M}} = c_M' + jc_M'' \\ c_M' \approx \sqrt{\dfrac{G'}{\rho_M}} \\ c_M'' \approx \sqrt{\dfrac{G'}{\rho_M}} \dfrac{1}{2Q_{mM}} \\ k_M = k_M' - jk_M'' = \dfrac{\omega_M}{c_M'} - j\dfrac{1}{2Q_{mM}}\dfrac{\omega_M}{c_M'} \end{cases} \qquad (19)$$

Here $k_M$ is the complex wavenumber whose real part determines the speed of sound, and the imaginary part determines the attenuation of the amplitude. $c_M$, $\rho_M$, G are the shear wave velocity, mass density and shear modulus of the testing specimen, respectively. $Q_{mM}$ is the quality factor of the specimen.

Similar as the derivation process of the equivalent circuit mode for the piezoelectric torsional transducer, based on the boundary conditions of the cylindrical specimen's velocity:

$$\begin{cases} \dot{\Theta}_M \big|_{z'=0} = -U_2 \\ \dot{\Theta}_M \big|_{z'=L} = U_3 \end{cases} \qquad (20)$$

and the boundary conditions of the forces:

$$\begin{cases} -F_2 = \int r\sigma_{\theta z'} \big|_{z'=0} dA \\ -F_3 = \int r\sigma_{\theta z'} \big|_{z'=L} dA \end{cases} \qquad (21)$$

The transport equations of the cylindrical specimen can be obtained as:

$$\begin{cases} F_2 = -\left(Z_M j\tan\dfrac{k_M L}{2} + \dfrac{Z_M}{j\sin k_M L}\right)U_2 - \dfrac{Z_M}{j\sin k_M L}U_3 \\ F_3 = -\dfrac{Z_M}{j\sin k_M L}U_2 - \left(Z_M j\tan\dfrac{k_M L}{2} + \dfrac{Z_M}{j\sin k_M L}\right)U_3 \end{cases} \qquad (22)$$

where

$$Z_M = \dfrac{G\pi D^4}{32 c_M} \qquad (23)$$

Taking into account the displacement continuity on the transducer-specimen interface, i.e.,

$$\Theta_P(z,t)\big|_{z=h,t=0} = \Theta_M(z',t)\big|_{z'=0,t=0} \qquad (24)$$

We can get $\omega_M = \omega_P$。

Based on Eq.(22) and the equivalent circuit model in Fig.2(a), the equivalent circuit model for the transducer-specimen composite system can be derived straightforwardly as shown in Fig.5(a). Taking into account the stress-free conditions on both ends, the equivalent circuit model can then be simplified to be Fig.5(b).

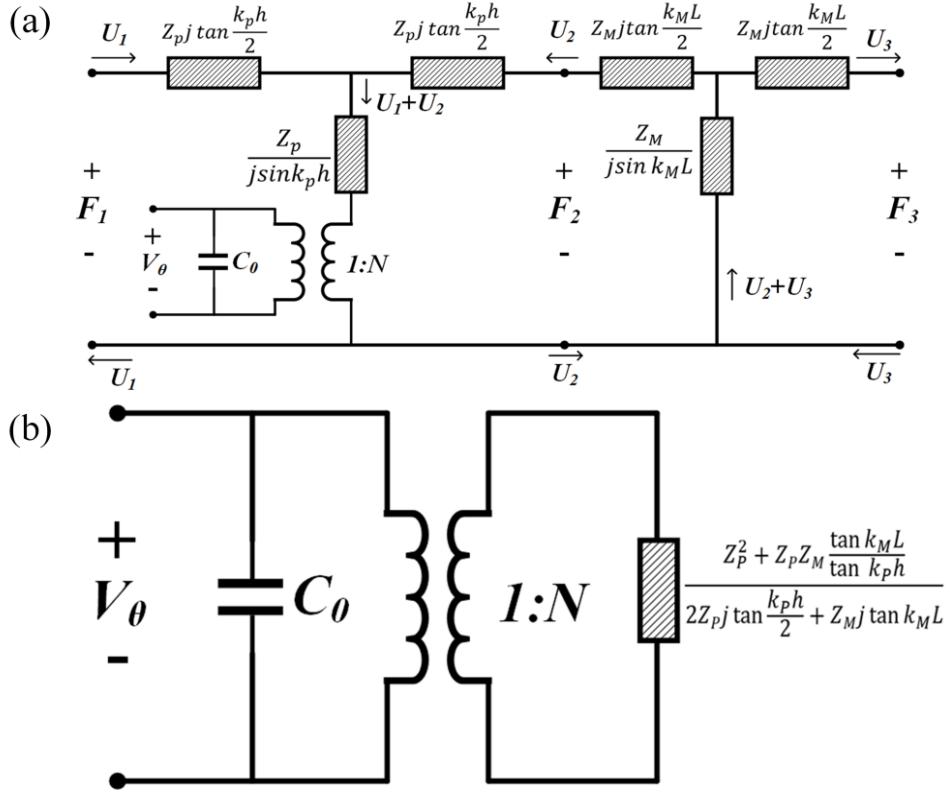

Fig.5  The equivalent circuit of the transducer-specimen composite system. (a) General case; (b) The case with the up/bottom faces stress-free.

From Fig.5b, the admittance of the transducer-specimen system can be obtained as follows:

$$Y = jC_0\omega + \frac{N^2}{Z} \qquad (25)$$

where

$$Z = \frac{Z_P^2 + Z_P Z_M \dfrac{\tan k_M L}{\tan k_p h}}{2Z_P j \tan \dfrac{k_p h}{2} + Z_M j \tan k_M L} \qquad (26)$$

### 2.3.2 Experimental validation

To verify the equivalent circuit model for the transducer-specimen composite system, the frequency-dependent admittance curves of the torsional transducer bonded on three different specimens (1045 steel bars with the length of 55mm and 100mm, PMMA bar with the length of 55mm) were measured using an impedance analyzer (Agilent 4294A) and compared with the theoretical results based on Eq.(25), as shown in Fig.6. In the theoretical calculations, the employed material parameters and specimen size were listed in Table I.

It can be seen from Fig.6 that the for all the three testing specimens, the theoretical admittance curves can be very close to the measured admittance curves if using the suitable internal friction ($Q_{mM}^{-1}$) for the specimen, which indicates the validity of the equivalent circuit model. From Fig.6, it can also be seen that for longer specimen (the 100mm-long 1045 steel bar), the first torsional resonance is almost invisible. This is because that for longer specimens, the first torsional resonance frequency is much lower than that of the piezoelectric torsional transducer, making the vibration amplitude of the composite system rather small thus the first resonance peak is not so significant. For longer specimens, the higher resonance modes can be used to derive the shear modulus.

Table I. Material parameters and specimen sizes of the transducer-specimen system used in the equivalent circuit model verification.

| PZT torsional transducer | | 1045 steel bar | |
|---|---|---|---|
| $h$ | 1.92mm | L | 55.00mm;100.00mm |
| d | 5.40mm | D | 12.00mm |
| D | 12.00mm | $\rho_M$ | 7780kg/m$^3$ |
| $\rho_P$ | 7500kg/m$^3$ | G | 82.90Gpa |
| $s_{44}^{\prime E}$ | $43.5 \times 10^{-12}$m$^2$/N | | |
| $P_{11}^{\prime \sigma}$ | 3130$\varepsilon_0$ | PMMA bar | |
| $d_{15}$ | $741 \times 10^{-12}$C/N | L | 55.00mm |
| $\varepsilon_0$ | $8.8542 \times 10^{-12}$ | D | 12.00mm |
| $Q_{mP}$ | 18 | $\rho_M$ | 1195kg/m$^3$ |
| $Q_e$ | 50 | G | 2.07Gpa |

It can also be seen from Fig.6 that with the increasing internal friction, the resonance peak becomes less significant. For the PMMA specimen whose internal friction is very large ($Q_{mM}^{-1} \sim 0.04$), all the resonance peaks are almost invisible, as shown in Fig.6(c).

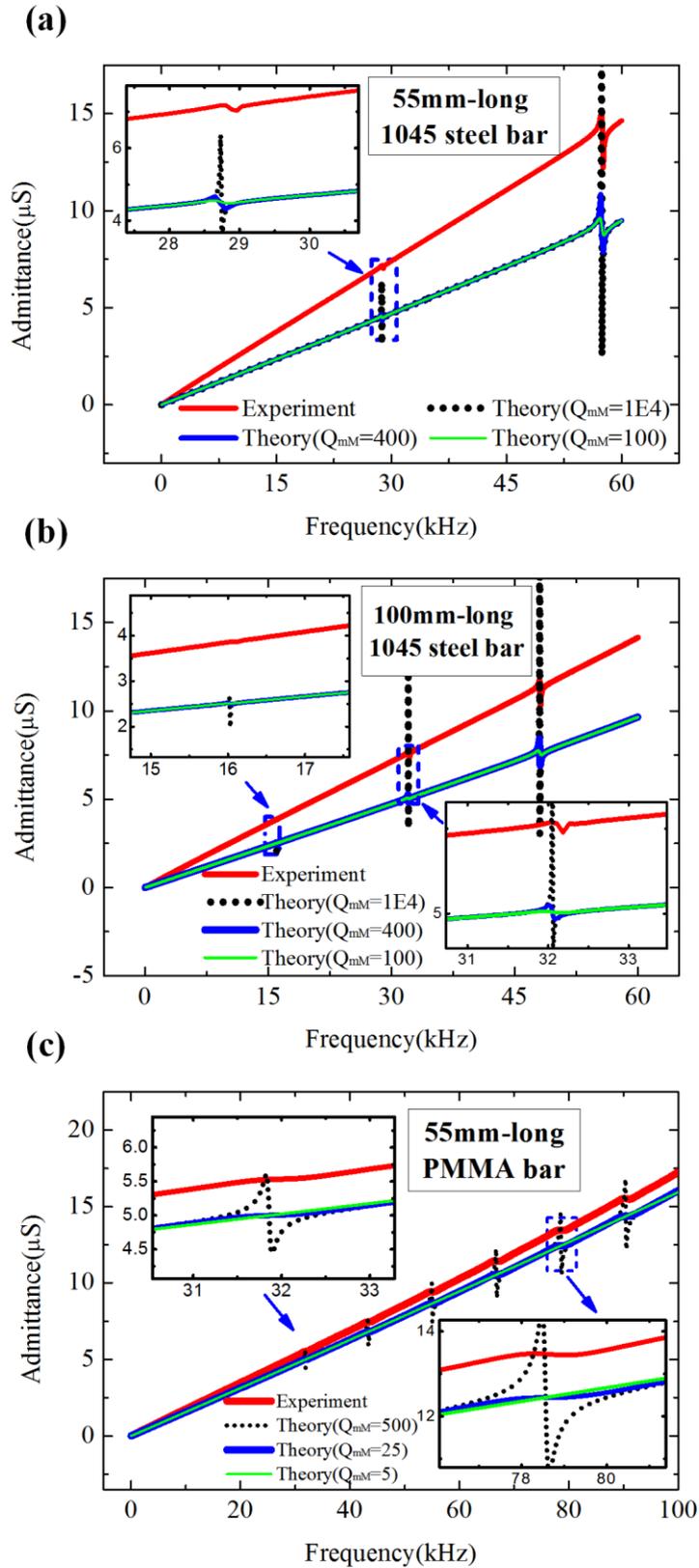

Fig.6 Experimental and theoretical admittance curves of the torsional transducer bonded on different cylindrical specimens with different prescribed internal friction. (a) 55mm-long 1045 steel bar; (b) 100mm-long 1045 steel bar; (c) 55mm-long PMMA bar.

### 2.3.3 Extraction of the shear modulus and internal friction

**a) Shear modulus**

From Fig.6, we see that the internal friction of the specimen has little effect on the resonance frequency. Thus, to determine the specimen's shear modulus using the resonance frequency, internal friction can be neglected and all parameters are real numbers at this time. Thus based on Eq.( 25), when the admittance gets its maxima, the impedance is zero, i.e.,

$$Z_P + Z_M \frac{\tan k_M L}{\tan k_P h} = 0 \tag{27}$$

By substituting the expressions for $Z_P$ and $Z_M$ into Eq.(27), the relationship between the first resonance frequency $f_{r1}$ and the specimen's shear modulus $G$ can be obtained as:

$$\tan\left(2\pi f_{r1} \sqrt{\frac{\rho_M}{G}} L\right) = -2\left(1 - \frac{d^4}{D^4}\right) \rho_P h f_{P1} \sqrt{\frac{1}{\rho_M G}} \tan\left(\frac{\pi f_{r1}}{f_{P1}}\right) \tag{28}$$

Where $f_{P1}$ is the first resonance frequency of PZT transducer which can be measured by the impedance analyzer directly.

Note that Eq.(28) is a transcendental equation. With the measured first resonance frequency $f_{r1}$, the shear modulus $G$ can be obtained with high accuracy by numerical methods.

However, in practical applications, it is not convenient to solve Eq.(25) numerically and an explicit expression for the shear modulus is requested. Bearing in mind that usually $L \gg h$, when the transducer-specimen system approaches its first resonance, for most specimens we have

$$\begin{cases} \tan\left(2\pi f_{r1} \sqrt{\frac{\rho_M}{G}} L\right) \approx 2\pi f_{r1} \sqrt{\frac{\rho_M}{G}} L - \pi \\ \tan\left(\frac{\pi f_{r1}}{f_{P1}}\right) \approx \frac{\pi f_{r1}}{f_{P1}} \end{cases} \tag{29}$$

Then Eq.(28) can be simplified to be

$$G = \frac{\left[\rho_M L + \rho_p h\left(1-\frac{d^4}{D^4}\right)\right]^2}{\rho_M} 4 f_{r1}^{\,2} \tag{30}$$

In the case that the rotary inertia of the transducer is at least two orders smaller than that of the specimen (i.e., $\frac{\rho_p h(1-\frac{d^4}{D^4})}{\rho_M L} < 0.01$), Eq.(30) can be further reduced to be:

$$G = 4\rho_M L^2 f_{r1}^{\,2} \tag{31}$$

That is, the influence of the piezoelectric transducer on the resonance frequency of the composite system can be neglected in this case with the induced testing error less than 2%.

**b) Internal friction**

Since in Fig.6, the sharpness of the resonance peak of the admittance curves varied continuously with the varied internal friction ($Q_{mM}^{-1}$), $Q_{mM}^{-1}$ can be extracted by fitting the theoretical admittance curve to the measured curve. When the internal friction of the testing sample is not very large, it can be approximately obtained by using the following formula which was derived following the early work by Zacharias:[24]

$$\begin{cases} \dfrac{1}{Q_{mM}} = 2\left[1 - \dfrac{\rho_p h}{\rho_M L}(1-\dfrac{d^4}{D^4})\right]\dfrac{\sqrt{(f_{rn}-f_{0n})(f_{0n}-f_{an})}}{f_{0n}} \\ f_{0n} = f_{rn} + \dfrac{Y_{an}}{Y_{rn}+Y_{an}}(f_{an}-f_{rn}) \end{cases} \tag{32}$$

Where $f_{rn}$ is the resonance frequency, $f_{an}$ is the anti-resonance frequency. $Y_{rn}$ is the admittance value corresponding to the resonance frequency and $Y_{an}$ is the admittance value corresponding to the anti-resonance frequency.

**2.3.4 Repeatability of the torsional resonance method**

Now let us discuss the testing errors of the electromechanical resonance method for shear modulus measurement. As indicated above, Eq.(28) can be used to calculate the shear modulus accurately using the measured fundamental resonance frequency $f_{r1}$. The total testing errors can then be classified into three types here: i) system errors which are caused by the difference between the equivalent circuit model and the real transducer-specimen system, such errors are inevitable and difficult to estimate, but they can be reduced by calibration; ii) Numerical errors in solving Eq.(28), these errors can be well controlled by using advanced numerical methods and thus regarded to be negligible; iii) transferred errors from the measurement of other quantities, including $\rho_M$, $\rho_P$, $f_{P1}$, $D$, $d$, $L$, $h$, which can be well estimated and controlled. Therefore, the absolute testing errors on shear modulus measurement using this torsional resonance method can be estimated if a standard testing specimen with known shear modulus is used. However, in practice, it is difficult to get such a standard specimen with the exact size and thus the absolute testing errors were not examined in this work.

On the other hand, the repeatable errors can be easily evaluated and good repeatability is especially useful for a practical testing method. Here the piezoelectric torsional transducer is bonded onto the specimen using the 502 epoxy cement, which can be easily removed by using acetone. To estimate the repeatable errors, for the first testing, the transducer was bonded onto the specimen and the specimen's shear modulus was measured, then the transducer was removed from the specimen. For the second testing, the transducer was bonded onto the same specimen again and the shear modulus measurement was repeated. Successive testing can be done similarly. So the difference between different testing on a same specimen is mainly due to the different bonding conditions. In this work, to reduce the repeatable errors, the thickness of the bonding epoxy in each testing is to be controlled as the same as possible.

Fig.7 shows that typical fundamental torsional resonance curves of the transducer bonded on a 100mm-long steel bar and a 100mm-long aluminum bar for two measurements. It can be seen that for both specimens, the repeatability errors for the torsional resonance frequency measurement is typically less than 0.1%. According to Eq.(30), the repeatability errors for the shear modulus measurement can be within 0.2%, which is good enough to sense the modulus variations during phase transitions[8-10] or material degradations.[25]

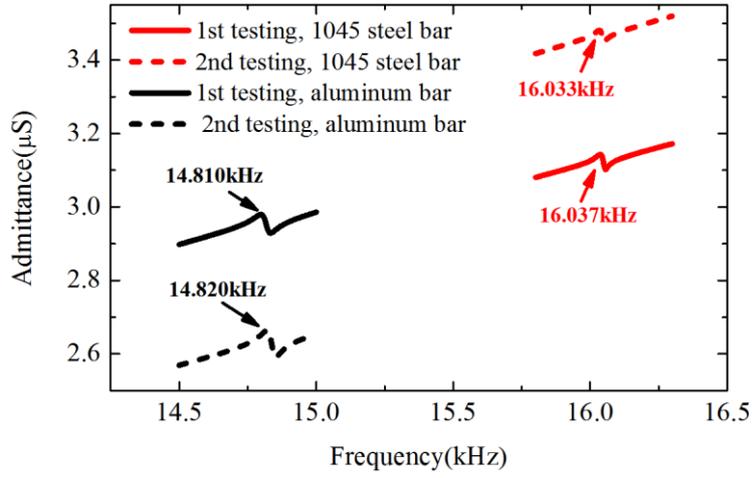

Fig.7 Typical testing repeatability for the fundamental torsional resonance of the transducer-specimen (100mm long) composite system.

**2.4 Shear modulus and internal friction measurement based on torsional wave propagation**

**2.4.1 Shear modulus**

Since the two-half-ring based piezoelectric torsional transducer is bonded onto the cylindrical specimen, the shear modulus of the specimen can also be measured based on torsional wave propagation. The piezoelectric torsional transducer is expected to excite torsional guided waves in the cylindrical specimen among which the fundamental torsional mode T(0,1) should be dominant.[26] This is because the deformation of the torsional transducer is uniform along the circumferential direction. Since T(0,1) mode is non-dispersive and its velocity always equals that of the bulk shear wave. The shear modulus can then be obtained via the following formula if the shear wave velocity is measured:

$$v_s = \sqrt{\frac{G}{\rho}} \qquad (33)$$

Where the density $\rho$ can be obtained by measuring the mass, the diameter and the length of the cylindrical specimen.

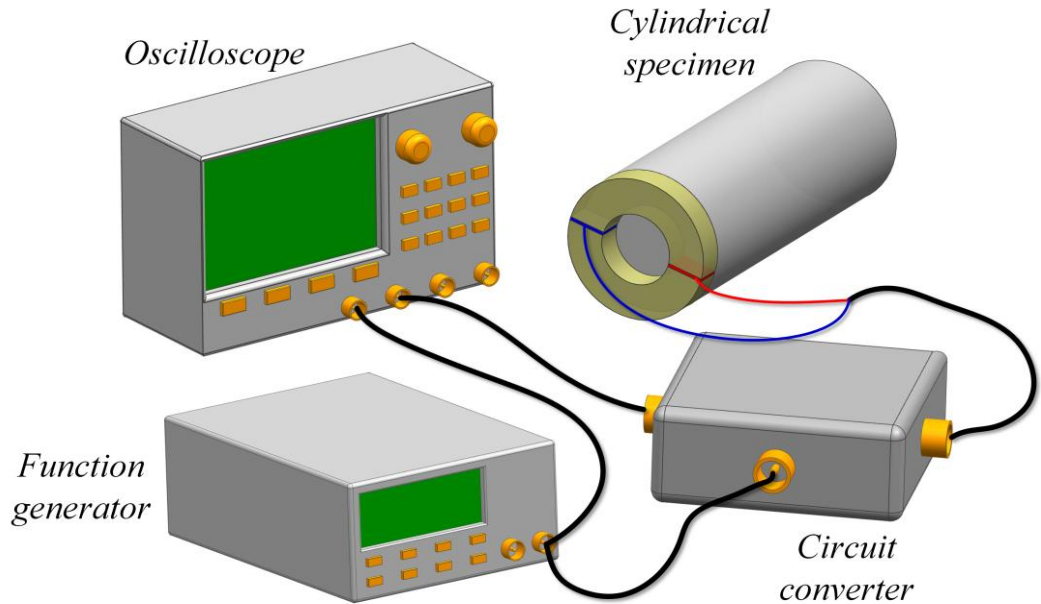

Fig.8 Experimental setup for shear modulus and internal friction measurement based on torsional wave propagation

In this work, the group velocity (or time of flight) of the T(0,1) wave mode is measured by using the conventional pulse-echo method. Fig.8 shows the experimental setup for the measurement of time of flight. A five-cycle or seven-cycle Hanning window-modulated sinusoid signal generated by a functional generator (Agilent 33220A) was used to drive the torsional transducer to generate torsional wave in the specimen. A circuit converter was used to make the torsional transducer both serve as wave actuator and sensor. Both the drive signal and the received signals were recorded and displayed by the oscilloscope. Note that the wave travel time inside the transducer was subtracted when measuring the time of flight in the specimen.

It should be noted that the torsional transducer is usually excited around (or at least not far from) its own resonance frequency (about 460kHz in this work) to ensure that enough energy can be transferred to the testing specimen. Typically, the resonance frequency of the torsional transducer is well above the cutting off frequency of the second (or higher) order torsional wave T(0,2) in a cylindrical sample. For example, Fig.9 shows the phase velocity (up, solid line) and group velocity (bottom, dashed line) curves of different torsional wave modes in a 12-mm diameter 1045 steel specimen. It can be seen that the torsional transducer's resonance frequency of about 460kHz is even higher than the cutting off frequency of the third order torsional wave mode T(0,3) in the steel specimen. Thus, T(0,2) and T(0,3) modes may also be excited in this case and signal processing should be used to extract the T(0,1) mode.

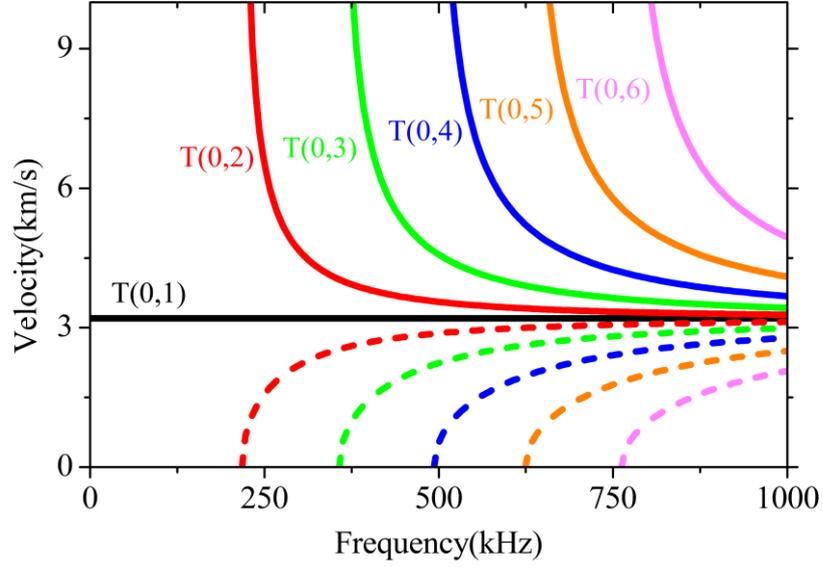

Fig.9 The phase velocity (solid line) and group velocity (dashed line) versus frequency curves for a 12mm-diameter cylindrical steel sample of infinite length

**2.4.2 Internal friction**

According to Eq.(19), when the internal friction is taken into account, the wavenumber is a complex number, and the harmonic wave in the specimen can be written as:

$$A_0 e^{jk(x-v_s t)} = A_0 e^{k''(x-v_s t)} e^{jk'(x-v_s t)} \tag{34}$$

Then the logarithmic attenuation $\zeta$ of the torsional wave is calculated to be

$$\zeta = \ln\left(\frac{A_0 e^{k''(x-v_s t)}\big|_{x=0,t=0}}{A_0 e^{k''(x-v_s t)}\big|_{x=0,t=2L/v_s}}\right) = 2Lk'' \tag{35}$$

The internal friction $Q_{mM}^{-1}$ is expressed as:

$$\frac{1}{Q_{mM}} = \frac{\zeta}{L}\frac{v_s}{2\pi f} \tag{36}$$

## 3. Measurement results and discussions

In this work, the shear modulus and internal friction of four types of materials, i.e., 1045 steel, aluminum, quartz glass and PMMA, were measured using both the torsional resonance method and the torsional wave propagation method. The testing results were presented in Section 3.1 and 3.2, respectively.

### 3.1 Measurement results using the torsional resonance method

Table II   Shear modulus and internal friction measurement results of four types of materials using the torsional resonance method

| Testing materials | density ($Kg/m^3$) | Length (mm) | First torsional resonance (kHz) | Shear modulus (GPa) based on | | | Reference shear modulus (GPa) | measured $Q_{mM}^{-1}$ ($10^{-3}$) | Reference $Q_{mM}^{-1}$ ($10^{-3}$) |
|---|---|---|---|---|---|---|---|---|---|
| | | | | Eq.(28) | Eq.(30) | Eq.(31) | | | |
| 1045 steel | 7780 ± 7.8 | 100 ± 0.05 | 16.03 ± 0.01 | 82.93 ± 0.19 | 82.83 ± 0.19 | 79.96 ± 0.19 | 78.8 ~ 83.8 | 2.34 | 0.5 ~ 4.0 |
| Aluminum | 2773 ± 2.8 | 100 ± 0.05 | 14.81 ± 0.01 | 26.88 ± 0.06 | 26.81 ± 0.06 | 24.33 ± 0.06 | 25.4 ~ 27.1 | 5.20 | 1.5 ~ 7.0 |
| Quartz glass | 2193 ± 2.2 | 100 ± 0.05 | 17.72 ± 0.01 | 31.21 ± 0.06 | 31.12 ± 0.06 | 27.54 ± 0.06 | 30.0 ~ 33.2 | 9.79 | 5.0 ~ 20.0 |
| PMMA | 1195 ± 1.2 | 55 ± 0.05 | 9.98 ± 0.01 | 2.06 ± 0.005 | 2.11 ± 0.005 | 2.62 ± 0.005 | 2.0 ~ 2.3 | 12.4 | 20.0 ~ 100.0 |

The measured shear modulus and internal friction ($Q_{mM}^{-1}$) of the four types of materials using the torsional resonance method were listed in Table II where the reference values from literatures were also presented for comparison. It can be seen from Table II that the calculated shear modulus G based on the exact solution in Eq.(28) is very close to that based on the approximate solution in Eq.(30), with the deviations of less than 0.3% for the first three lossless materials and of about 2% for the lossy PMMA. Furthermore, all the measured shear modulus based on Eq.(28) and Eq.(30) fit well with the reference values from literatures. Bearing in mind that Eq.(30) is an explicit expression for the shear modulus, it is more convenient for practical use than the implicit Eq.(28). In comparison, the calculated shear modulus based on Eq.(31), i.e., neglecting the transducer's effect, differs up to 10% from the value based on Eq.(30) for the three lossless materials and about 30% for the lossy PMMA. Thus, the transducer's effect cannot be neglected for an accurate measurement. Therefore, Eq.(30) is the best solution to calculate the shear modulus based on the measured torsional resonance frequency.

From Table II, it can also be seen that the measured internal friction ($Q_{mM}^{-1}$) using the approximate formula Eq.(32) are also close to the reference values for the three types of lossless materials. While for the lossy PMMA, the calculated internal friction is obviously smaller than the reference values. This may be due to the fact that the approximate Eq.(32), which is based on the small damping assumption, cannot be applicable to the lossy materials. Thus, the torsional resonance method is not suitable for internal friction measurement on lossy materials. In addition, since the internal friction measured in this work is for the torsional mode, and the reference values are based on the tension/compression mode or bending mode, it is normal that they differ from each other to some extent.

### 3.2 Measurement results using the torsional wave propagation method

Shear modulus and internal friction measurement were also conducted on the four types of materials based on the torsional wave propagation method using 5-cycle or 7-cycle Hanning windowed sinusoidal signal with the central frequency of 150kHz. Fig.10 shows the original received wave signals and that after continuous wavelet transformation (CWT) in a 200mm-long 1045 steel bar and a 100mm-long PMMA bar, respectively. As shown in Fig.10(a) and 10(b), the torsional wave velocity can be determined by measuring the time of flight between the two successive received wave package. The wave traveling time inside the 2mm-thick transducer should be removed using the shear wave velocity of 1750 m/s. The internal friction can also be obtained by measuring the amplitude of the two successive wave package and using Eq.(36). However, it should be noted that this approach works very well for the lossless materials, such as the steel, aluminum, quartz glass measured in this work. For lossy materials, such as the PMMA bar measured in this work, the second echo is almost invisible, as shown in Fig.10(c) and 10(d). Therefore, to measure the internal friction of lossy materials using the wave propagation method, short specimens are preferred to and large testing signals are required.

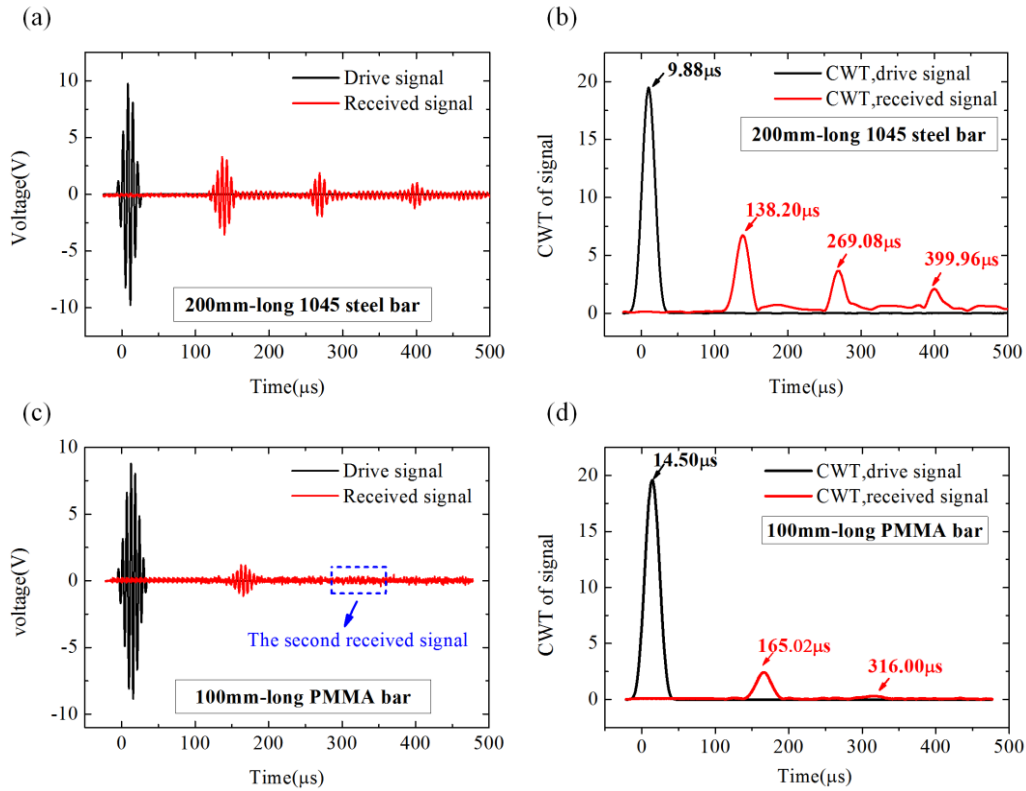

Fig.10 Drive and received torsional wave signals for 200mm-long 1045 steel (up) and 100mm-long PMMA (bottom) at 150kHz. Left: original signals; right: signals after continuous wavelet transformation (CWT).

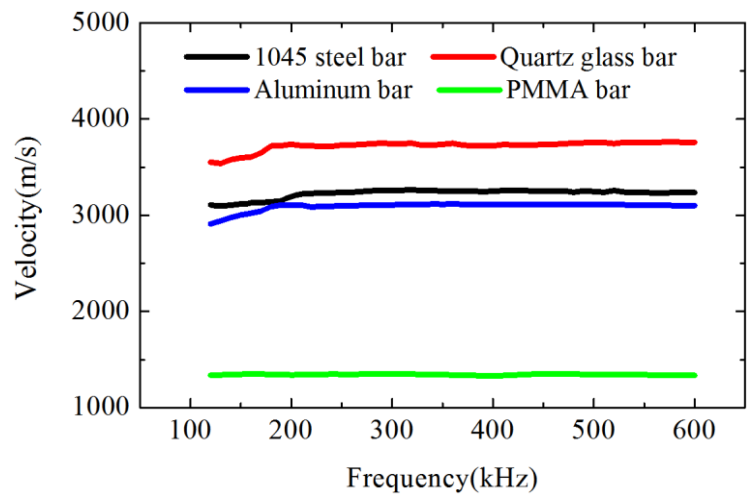

Fig.11 Frequency dependent shear velocity measurement results for four types of materials using the torsional wave propagation method.

It should be noted that the received wave signals in Fig.10 are frequency dependent. Fig.11 shows the measured frequency dependent velocities for the four types of materials. It can be seen that for the three lossless materials, i.e., 1045 steel, aluminum, and quartz glass, the measured torsional wave velocity slightly increased with the testing frequency below 200kHz. This is partially due to the fact that at low frequencies, the time of flight between the reflected signals cannot be accurately determined because of the large wave length at low frequencies (about 22mm for the 1045 steel at 150kHz, more than 1/10 of the specimen length). When the testing frequency is above 200kHz, the measured torsional wave velocity turns to be stable for all the four types of materials. Therefore, when measuring the torsional wave velocity of the cylindrical specimen, typically high frequency signal is required. In our experiences, the testing wavelength should be smaller than 1/12 of the specimen length.

The measured stabilized shear wave velocity, shear modulus and internal friction for the four types of materials are listed in Table III. It can be seen that for all the four types of materials, the measured shear modulus is very close to that measured by using the torsional resonance method, and also consistent with the reference values, which indicates the validity of both methods in shear modulus measurement. As to the internal friction measurement, for the three types of lossless materials, the measurement results using the wave propagation method are also very close to that by the torsional resonance method. The measured internal friction of PMMA, which is 30.7E-3, is also consistent with the reference values of (20~100)E-3. As mentioned before, the approximate Eq.(32) is not suitable for internal friction measurement on very lossy materials based on the resonance method. Thus, for shear mode internal friction measurement on lossy materials, the torsional wave propagation method seems to be the unique solution if the torsion pendulum method cannot be applicable.

Table III. Shear modulus and internal friction measurement results of four types of materials using the torsional wave propagation method.

| Testing materials | Density ($Kg/m^3$) | Length (mm) | Shear wave velocity $v_s(m/s)$ | Shear modulus(GPa) | Internal friction ($10^{-3}$) |
|---|---|---|---|---|---|
| 1045 steel | 7780 ± 7.8 | 200 ± 0.05 | 3242.54 ± 20.00 | 81.80 ± 1.15 | 2.38 |
| Aluminum | 2773 ± 2.8 | 100 ± 0.05 | 3107.04 ± 30.00 | 26.77 ± 0.50 | 5.47 |
| Quartz glass | 2193 ± 2.2 | 100 ± 0.05 | 3740.42 ± 23.00 | 30.68 ± 0.40 | 8.87 |
| PMMA | 1195 ± 1.2 | 100 ± 0.05 | 1341.92 ± 10.00 | 2.15 ± 0.03 | 30.7 |

**3.3 Discussions**

From the measurement results in Table II and Table III and the measurement principles in Section 2, it can be seen that for shear modulus measurement, if the specimen is not very long (say less than 100mm), the torsional resonance method is better than the torsional wave propagation method for its easy use and very good repeatability. The torsional resonance method can also be applicable for shear modulus measurement on long specimens using high order resonances, while in that case, extra efforts should be taken to identify the resonance order and the shear modulus calculation formula would be more complicated. So for long specimens, the wave propagation method is suggested.

With regard to the internal friction measurement, as discussed above, the resonance method cannot be applicable to lossy materials which can be measured using the wave propagation method. It should be noted that although both the resonance method and wave propagation method work well for the internal friction measurement on lossless materials, very accurate measurement on internal friction is not possible because the interface effect between the transducer and the specimen cannot be exactly modelled. For the lossless metallic materials, the internal friction can be accurately measured based on the contactless electromagnetic acoustic resonance method (EMAR).[27] While the EMAR is not applicable to non-metallic materials whose internal friction can be accurately measured based on the wave propagation method using very long specimens.

## 4. Conclusions

In summary, we proposed two shear modulus measurement methods based on torsional resonance and torsional wave propagation using a same piezoelectric torsional transducer. The shear modulus measurement principles for both methods were derived and that for internal friction measurement were also presented. Testing results show that the torsional resonance method is more suitable for shear modulus measurement on short or middle-length bars (below 100mm) with high repeatability better than 0.2%. For longer specimen, the torsional wave propagation method is more suitable. As to the internal friction measurement, both methods works well for the lossless materials while for the lossy materials, only the wave propagation method is applicable. The proposed two methods using only one torsional transducer provide a quick and reliable solution to shear modulus and internal friction measurement, which can be widely used in near future.

**Acknowledgement**

This work is supported by the National Natural Science Foundation of China under Grant Nos.11672003.